\documentclass[12pt]{article}
\usepackage[english]{babel}
\usepackage[utf8]{inputenc}
\usepackage[T1]{fontenc}

\def\mF{\mathcal{F}}
\def\bP{\mathbf{P}}
\usepackage{amsmath}
\def\bC{\mathbf{C}}

\def\mC{\mathcal{C}}
\usepackage{amsfonts}

\usepackage{url}
\usepackage[dvips]{graphicx}
\usepackage{calc}
\usepackage{subfigure}
\usepackage{multirow}

\def\mL{\mathcal{L}}
\def\mH{\mathcal{H}}
\def\mD{\mathcal{D}}

\def\bx{\mathbf{x}}
\def\tmH{\tilde{\mH}}
\newcommand{\pb}[1]{\left\{#1\right\}}

\def\bA{\mathbf{A}}

\def\mR{\mathcal{R}}

\def\mG{\mathcal{G}}
\def\mS{\mathcal{S}}
\def\bG{\mathbf{G}}
\def\ba{\mathbf{a}}
\def\by{\mathbf{y}}

\def\bT{\mathbf{T}}

\def\mM{\mathcal{M}}
\def\tmH{\tilde{\mH}}
\def\bH{\mathbf{H}}
\def\bM{\mathbf{M}}

\begin{document}
	\begin{titlepage}
	\begin{center}
		{\Large{ \bf Brown-Kucha\v{r} Mechanism for Unstable D-Brane
		at Large Tachyon Regime}}
		
		\vspace{1em}  
		
		\vspace{1em} J. Kluso\v{n} 			
		\footnote{Email addresses:
			klu@physics.muni.cz (J.
			Kluso\v{n}) }\\
		\vspace{1em}
		\textit{Department of Theoretical Physics and
			Astrophysics, Faculty of Science,\\
			Masaryk University, Kotl\'a\v{r}sk\'a 2, 611 37, Brno, Czech Republic}
		
		\vskip 0.8cm
		
		%
		%
		%
		%
		%
		%
		
		\vskip 0.8cm
		
	\end{center}

\begin{abstract}
We study space-time filling unstable D-brane coupled to gravity. We find Hamiltonian for this D-brane
and then we implement Brown-Kucha\v{r} formalism for this system. We calculate algebra of constraints of deparametrized
theory and we show that for large $T$ the Poisson brackets between new constraints are zero. Then 
we define gauge invariant functions on phase space following deparametrization construction introduced in  \cite{Thiemann:2006up}.
\end{abstract}

\bigskip

\end{titlepage}

\newpage

\section{Introduction and Summary}\label{first}
According to Dirac \cite{Dirac} all observables have to be constant along gauge orbits and therefore have vanishing Poisson brackets with all first class   constraints that are generators of gauge transformations. In case of General Relativity it can be shown that the Hamiltonian is given as sum of the first class constraints and hence Hamiltonian vanishes on the constraint surface \cite{Dirac:1958sc,Arnowitt:1962hi}. Then it is clear that all proper observables do not evolve dynamically which is known as problem of time. 
The fact that there is no evolution with respect to the time is in conflict with our everyday experience where
we experience gravitational observation as dynamical process. On the other hand the canonical Hamiltonian 
describes evolution with respect to coordinate time which really does not have physical meaning thanks to  the manifest
diffeomorphism invariance of any theory coupled to gravity. What really makes sense is evolution with respect
to other fields. 

Nice proposal how to solve problem of time was suggested by T. Thiemann in \cite{Thiemann:2006up} that is based on fundamental work \cite{Brown:1994py}. The main idea of this proposal is to perform deparametrization of theory which means that we find new Hamiltonian-like  constraint in the form $
\mC=\pi+H\approx 0$ where $\pi$ is momentum conjugate to scalar field $\phi$ and where $H$ is positive function of remaining phase space degrees of freedom which does not depend on $\pi$ and $\phi$. Then it was shown in
\cite{Thiemann:2006up} that it is possible to construct physical observables and Hamiltonian $\bH=\int d^3\bx H$ that generates their time evolution. In this approach the scalar field  can be interpreted as physical clock. Even if  this scalar field is not directly observable 
 its presence has an important consequence \cite{Thiemann:2006up}. First of all physical Hamiltonian is slightly different from the usual one that could have cosmological implications as was also shown 
 in \cite{Thiemann:2006up} . Further, the true time evolution is generated by physical Hamiltonian $\bH$ rather than Hamiltonian constraint which also solves famous problem of time.

In this paper we would like to implement similar approach to the situation when the scalar field that serves for deparametrization of gravity, corresponds to the tachyon field living on the world-volume of non-BPS D9-brane. 
\cite{Sen:1999md,Garousi:2000tr,Bergshoeff:2000dq,Kluson:2000iy}
\footnote{Non-BPS D9-brane is unstable $9+1$ dimensional object in superstring Type IIA theory which is defined in $10-$dimensional space-time. For that reason we consider deparametrization of ten dimensional gravity.}. With the help of the tachyon field and an action for non-BPS D9-brane we can perform deparametrization of General Relativity  in the regime of  large tachyon field so that tachyon is directly related to the time evolution.  We should stress that the idea that  open string tachyon could be related to the physical time was suggested by A. Sen in  seminal paper \cite{Sen:2002qa} and recently in  \cite{Sen:2023qya}. In more details, it was shown there that for large tachyon field the tachyon can be interpreted as  non-rotating,
non-interacting dust. Further, Brown-Kucha\v{r} mechanism was also applied for unstable D-brane that is function of tachyon only \cite{Sen:2002qa}. In this paper we extend this analysis to the general form of non-BPS D9-brane action including gauge fields and possible other scalar fields in order to see how the presence of other fields in the action affects the Brown-Kucha\v{r} analysis for non-BPS D9-brane. 

Let us be more precise with the description of the current work. Since the goal is to study space-time filling D9-brane coupled to gravity we firstly determine  Hamiltonian for this object. We derive  this Hamiltonian from the Hamiltonian for reparametrization invariant D-brane action after fixing the static gauge. Then we show that  gravity coupled to non-BPS D9-brane is completely constrained system where the Hamiltonian is given as the sum of nine spatial diffeomorphism constraints and  Hamiltonian one. We also explicitly check that they are the first class constraints which is necessary condition for Brown-Kucha\v{r} formalism. Then we proceed to the deparametrization  formulation of General Relativity. We firstly solve 
spatial diffeomorphism constraints for partial derivative of tachyon 
and then we insert this result to the  Hamiltonian constraint. We further  solve the Hamiltonian  constraint for $p_T$ and we obtain new one which is linear in $p_T$ and where  the remaining part  will be crucial for the construction of physical Hamiltonian.  As the next step we explicitly check that the Poisson bracket between new constraint is zero for large tachyon field. The condition that the tachyon is large is very  important since then we get that the new constraint does not depend on $T$ explicitly. Note that we derive this result even in the case of
non-zero gauge field represented by non-zero electric flux which has nice physical interpretation as the gas of fundamental strings \cite{Sen:2003bc,Sen:2000kd,Yee:2004ec,Gibbons:2000hf}.

Finally we proceed to the construction of Dirac
 observables in deparametrized theory, following \cite{Thiemann:2006up}. We show that such observables Poisson commute with Hamiltonian and spatial diffeomorphism constraints where our proof is different from the analysis presented in \cite{Thiemann:2006up}. We also calculate derivative of this observable with respect to parameter $\tau$ and we show that it is equal to the Poisson bracket between physical Hamiltonian and this observable
 and it corresponds to true  evolution in deparametrized theory \cite{Thiemann:2006up}.

Let us outline our results and suggest possible extensions of this work. We firstly found Hamiltonian for space-time filling unstable D9-brane coupled to gravity and we showed that the Hamiltonian for whole system is given as the sum of the first class constraints. Then we apply Brown-Kucha\v{r} mechanism for this system and we got new constraint that is linear in momentum conjugate to the tachyon field.  We showed that the Poisson brackets between this new constraint vanishes in case of large tachyon field $T$ even with the presence of the  electric flux. Finally we construct true Dirac observables that Poisson commute with Hamiltonian and diffeomorphism constraints. 

This paper can be extended in many directions. The first one which is currently under study is  calculations of Poisson brackets between new deparametrized constraints for arbitrary value of the tachyon field and we would like to see whether they are also exactly zero or  they are proportional to the linear combinations of constraints. Further, we mean that it would be nice to see how Brown-Kucha\v{r} mechanism could be used for space-time filling 
D9-brane anti-D9-brane system in Type IIB theory
\cite{Sen:2003tm,Garousi:2004rd,Kluson:2022uws}. Finally it would be also interesting to study
how the presence of non-zero electric flux could affect cosmological evolution, 
 following analysis performed in 
\cite{Thiemann:2006up}.

This paper is organized as follows. In the next section (\ref{second}) we find Hamiltonian for space-time filling non-BPS D9-brane. We determine all constraints and show that they are the first class ones. In section (\ref{third}) we introduce
Brown-Kucha\v{r} mechanism for this system. We also calculate
Poisson brackets between constraints for large tachyon field. Finally in section (\ref{fourth}) we introduce Dirac observables 
and check that they Poisson commute with spatial diffeomorphism and Hamiltonian constraints. 

\section{Hamiltonian For Unstable D-Brane Coupled to Gravity }\label{second}
In this section we find Hamiltonian for space-time filling unstable D-brane which couples to gravity. 
The most elegant way how to find such Hamiltonian is to consider reparametrization invariant
action for unstable D-brane that has the form
\begin{equation}\label{Snongen}
	S=-\int d^{p+1}\sigma V(T)\sqrt{-\det \bA_{\alpha\beta}} \ , 
\end{equation}
where
\begin{equation}
	\bA_{\alpha\beta}=g_{MN}\partial_\alpha x^M\partial_\beta x^N+
	\lambda F_{\alpha\beta}+M_{IJ}\partial_\alpha \phi^I\partial_\beta \phi^J+\lambda\partial_\alpha T\partial_\beta T \ , 
\end{equation}
where $\sigma^\alpha,\alpha,\beta=0,1\dots,p$ label world-volume directions of Dp-brane so that $\partial_\alpha\equiv \frac{\partial}{\partial \sigma^\alpha}$. Further, 
$x^M.M.N=0,1,\dots 9$ parameterize embedding of Dp-brane into $10-$dimensional target space-time,
\footnote{Our construction is based on space-time filling unstable D9-brane that exists in Type IIA string theory. In case of Type IIB theory we should instead consider D9-brane anti-D9-brane pair. We expect that qualitative discussion will be the same as in case of non-BPS D9-brane analyzed in this paper.}	 
$\lambda=2\pi\alpha'=l_s^2$ where $l_s$ is string length,  and $F_{\alpha\beta}=\partial_\alpha A_\beta-\partial_\beta A_\alpha$, where 
$A_\alpha$ is a gauge field living on the world-volume of Dp-brane. For general purposes we also included
set of scalar fields $\phi^I,I=1,2,\dots,D$ with internal metric $M_{IJ}$ that depends on $\phi$ only. 
Finally $T$ is the tachyon field with the potential $V(T)$ with the property that $T$ has two stable
minima for $T_{min\pm}=\pm \infty$ where $V(T_{min \pm})=0$ while it has an unstable maximum at $T_{max}=0$ where $V(T_{max})=\tau_{nonBPS}$ where $\tau_{nonBPS}=\frac{\sqrt{2}2\pi}{(2\pi l_s)^{p+1}}$ is a tension of unstable Dp-brane. 

Now we are ready to find canonical form of the action (\ref{Snongen}) where we firstly determine corresponding momenta conjugate to $x^M,\phi_I,T$ and $A_\alpha$
\begin{eqnarray}\label{defM}
&&p_M=\frac{\partial \mL}{\partial (\partial_0 x^M)}=
-V(T)g_{MN}\partial_\beta x^N (\bA^{-1})_S^{\beta 0}\sqrt{-\det \bA} \ , \nonumber \\
&&p_I=\frac{\partial \mL}{\partial (\partial_0 \phi^I)}=
-V(T)M_{IJ}\partial_\beta \phi^J  (\bA^{-1})_S^{\beta 0}\sqrt{-\det \bA} \ , \nonumber \\
&&p_T=\frac{\partial \mL}{\partial (\partial_0 T)}=
-V(T)\lambda \partial_\beta T  (\bA^{-1})_S^{\beta 0}\sqrt{-\det \bA} \ , \nonumber \\
&&\pi^i=\frac{\partial \mL}{\partial (\partial_0 A_i)}=
-\lambda V(T)(\bA^{-1})^{i0}_A\sqrt{-\det\bA} \ , \quad \pi^0\approx 0 \ , \nonumber \\
\end{eqnarray}
where $(\bA^{-1})^{\alpha\beta}_{S,A}$ are defined as
\begin{equation}
(\bA^{-1})^{\alpha\beta}_S=(\bA^{-1})^{\alpha\beta}+
(\bA^{-1})^{\beta\alpha} \ , \quad 
(\bA^{-1})^{\alpha\beta}_A=(\bA^{-1})^{\alpha\beta}-
(\bA^{-1})^{\beta\alpha} \ . \quad 
\end{equation}
Now using following properties of matrices $\bA_{\alpha\beta}$ 
\begin{eqnarray}
&&	(\bA^{-1})^{\alpha\beta}_S \bA_{
		\beta\gamma}^S+(\bA^{-1})^{\alpha\beta}_A
	\bA_{\beta\gamma}^A=\delta^\alpha_\gamma \ , \nonumber \\
&&	(\bA^{-1})^{\alpha\beta}_A\bA_{
		\beta \gamma}^S+(\bA^{-1})^{\alpha\beta}_S \bA^A_{\beta\gamma}=0 \ , 
	\nonumber \\
&&	\bA^S_{\beta\gamma}(\bA^{-1})^{\gamma 0}_A+
	\bA^A_{\beta\gamma}(\bA^{-1})^{\gamma 0}_S=0 \nonumber \\
\end{eqnarray}
and also definition of momenta given in 
(\ref{defM}) we obtain following set of $p+1$ constraints 
\begin{eqnarray}
&&\mathcal{K}_i\approx   p_M\partial_i x^M+p_I\partial_i\phi^I+p_T\partial_i T+
 F_{ij}\pi^j \approx 0 \ , \quad i,j=1,\dots,p \ ,  \nonumber \\
&&\mathcal{K}=p_M g^{MN}p_N+p_IM^{IJ}p_J+\frac{1}{\lambda}p_T^2+
\frac{1}{\lambda^2}\pi^i\bA_{ij}^S\pi^j+V^2\det \bA_{ij}\approx 0
\nonumber \\
\end{eqnarray}
while the bare Hamiltonian is equal to
\begin{equation}
	H_B=\int d^p\sigma (p_M\partial_0 x^M+p_T\partial_0 T+
	p_I\partial_0 \phi^I+\partial_0 A_\alpha \pi^\alpha-\mL)=
	\int d^p\sigma \pi^i\partial_i A_0 \ . 
\end{equation}
It can be shown that $\mathcal{K}\approx 0, \mathcal{K}_i\approx 0 $ are first class constraints. Now we return to the case of space-time
filling D9-brane. It is convenient to impose the static
gauge which means that the space-time coordinates coincide with 
the world-volume  ones. Explicitly we have
\begin{equation}
x^M(\sigma^\beta)=\sigma^\alpha \delta_\alpha^M \ 
\end{equation}
that can be written in the form of gauge fixing functions 
\begin{equation}
	\mF^\alpha\equiv x^\alpha-\sigma^\alpha\approx 0 \ .
\end{equation}
It can be shown that $\mF^\alpha\approx 0$ are  second class constraints with $\mathcal{K}\approx 0 \ , 
\mathcal{K}_i\approx 0$ so that they strongly vanish and 
 can be solved explicitly. From $\mathcal{K}_i=0$ we express $p_i$ as
\begin{equation}
p_i=-p_I\partial_i\phi^I-p_T\partial_i T- F_{ij}\pi^j\equiv 
-\mH_i
\end{equation}
while from $\mathcal{K}=0$ we get
\begin{eqnarray}
&&p_0=\frac{-2g^{0i}p_i\pm \sqrt{4(g^{0i}p_i)^2-4g^{00}(\mD+p_ig^{ij}p_j)}}{2g^{00}}\ \ ,  
\nonumber \\
&&\mD=p_I M^{IJ}p_J+\frac{1}{\lambda}p_T^2+\frac{1}{\lambda^2}\pi^i\bA^S_{ij}\pi^j+V^2\det \bA_{ij} \ , \nonumber \\
&& \bA_{ij}=g_{ij}+M_{IJ}\partial_i\phi^I\partial_j \phi^J+\lambda\partial_i T\partial_jT+\lambda F_{ij} \ . 
\end{eqnarray}
At this place it is convenient to introduce $9+1$ formalism for the background gravity \footnote{For  review, see
	\cite{Gourgoulhon:2007ue}.}. We consider $10-$dimensional manifold
$\mathcal{M}$ with the coordinates $x^M \ , M=0,\dots,9$ and
where $x^M=(t,\bx) \ , \bx=(x^1,x^2,\dots,x^{9})$. We presume that this
space-time is endowed with the metric $g_{MN}(x^\rho)$
with signature $(-,+,\dots,+)$. Suppose that $ \mathcal{M}$ can be
foliated by a family of space-like surfaces $\Sigma_t$ defined by
$t=x^0=\mathrm{const}$. Let $h_{ij}, i,j=1,2,\dots,9$ denotes the metric on $\Sigma_t$
with inverse $h^{ij}$ so that $h_{ij}h^{jk}= \delta_i^k$. We further
introduce the operator $\nabla_i$ that is covariant derivative
defined with the metric $h_{ij}$.
We also define  the lapse
function $N=1/\sqrt{-g^{00}}$ and the shift function
$N^i=-g^{0i}/g^{00}$. In terms of these variables we
write  components of the metric $g_{MN}$ as
\begin{eqnarray}
	g_{00}=-N^2+N_i h^{ij}N_j \ , \quad g_{0i}=N_i \ , \quad
	g_{ij}=h_{ij} \ ,
	\nonumber \\
	g^{00}=-\frac{1}{N^2} \ , \quad g^{0i}=\frac{N^i}{N^2} \
	, \quad g^{ij}=h^{ij}-\frac{N^i N^j}{N^2} \ .
	\nonumber \\
\end{eqnarray}
Then using this form of the background metric we obtain that $p_0$ is equal to
\begin{eqnarray}
&&p_0=-N^i\mH_i - N\sqrt{\mD+\mH_ih^{ij}\mH_j} \ , \nonumber \\
\end{eqnarray}
where we have chosen $-$ sign in front of the square root in order to have positive Hamiltonian. 
In fact,  due to this gauge fixing we get that the action for D9-brane has the form
\begin{eqnarray}\label{gaugefixed}
&&	S_{D9}=\int d^{10} \sigma(p_I\partial_0 \phi^I+p_T\partial_0 T+p_M\partial_0 x^M+
	\pi^i\partial_0A_j-\Omega \mathcal{K}-\Omega^i\mathcal{K}_i-\pi^i\partial_iA_0)=
	\nonumber \\
&&=\int d^{10} x(p_I\partial_0 \phi^I+p_T\partial_0 T+p_0+
\pi^i\partial_0A_i+A_0\partial_i\pi^i) \ ,
\nonumber \\	
\end{eqnarray}
using the fact that for gauge fixed theory $\mathcal{K}$ and $\mathcal{K}_i$ strongly vanish. Then we see from (\ref{gaugefixed})
 that it is natural to identity $-p_0$ as Hamiltonian density for gauge fixed system. 
As a result the complete form of the action  for gravity together with the space-time filling unstable D9-brane has the form
\begin{eqnarray}
&&	S=S_{GR}+S_{D9}=\nonumber \\
&&=\int d^{10}x (\pi^{ij}\partial_0 h_{ij}+p_I\partial_0 \phi^I+
p_T\partial_0 T+\pi^i \partial_0 A_i-N\mC-N^i\mC_i+A^0\mathcal{E}) \ , \nonumber \\
\end{eqnarray}
where
\begin{eqnarray}
&&\mC=\frac{\kappa}{\sqrt{h}}(\pi^{ij}h_{ik}h_{jl}\pi^{kl}-
(\pi^{ij}h_{ij})^2)-\frac{1}{\kappa}\sqrt{h}r+
\sqrt{\mD+\mH_ih^{ij}\mH_j}\equiv \nonumber \\
&& \equiv \mH^G_\bot+\mC^{matt} \ , \nonumber \\
&&\mC_i=
-2\nabla_l\pi^{kl}h_{ki}+\mH_i \equiv \mH^G_i+\mH_i \equiv \tmH_i+p_T\partial_i T\approx 0 \ , \quad 
\mathcal{E}=\partial_i\pi^i \approx 0 \ , \nonumber \\
\end{eqnarray}
where we also used the fact that requirement of the preservation of the constraint $\pi^0\approx 0$ implies the constraint $\mathcal{E}\approx \partial_i\pi^i\approx 0$. Note that $\pi^{ij}$ are momenta conjugate to $h_{ij}$ and $r$ is scalar curvature calculated with the metric $h_{ij}$. Finally $\nabla_i$ is covariant derivative compatible with the metric components $h_{ij}$. Further,  the requirement of the preservation of the constraints 
$\pi_N\approx 0 \ , \pi_{N^i}\approx 0$ where $\pi_N$ and $\pi_{N^i}$ are momenta conjugate to $N$ and $N^i$ respectively implies an existence of the constraints 
\begin{equation}
	\mC\approx 0 \ ,  \quad  \mC_i\approx 0 \ . 
\end{equation}	
Before we proceed to the explicit form of Brown-Kucha\v{r} mechanism for this system we check that $\mC\approx 0 \ ,  \mC_i\approx 0$ are first class constraints. 
\subsection{Constraint Algebra}
In this section we explicitly determine algebra of constraints $\mC\approx 0$ and $\mC_i \approx 0$. 
It is convenient to  introduce their smeared forms
\begin{eqnarray}
&&	\bC(X)=\int d^9\bx X\mC\equiv \bC^G(X)+\bC(X)^{matt} \ , \nonumber \\
&&	\bC_S(Y^i)=\int d^9\bx Y^i\mC_i\equiv \bC_S^G(Y^i)+
	\bC_S^{matt}(Y^i) \ , \nonumber \\ 
\end{eqnarray}	
 where $X(\bx),Y(\by)$ are arbitrary smooth functions on $\Sigma$. Note that we also performed appropriate splitting of the constraints to gravitational and matter parts respectively. 

Let us now calculate Poisson brackets between constraints using canonical Poisson brackets 
\begin{eqnarray}
&&	\pb{h_{ij}(\bx),\pi^{kl}(\by)}=
\frac{1}{2}(\delta_i^k\delta_j^l+\delta_i^l\delta_j^k)
\delta(\bx-\by) \ , \quad 
\pb{A_\alpha(\bx),\pi^\beta(\by)}=\delta_\alpha^\beta\delta(\bx-\by) \ , \nonumber \\
&&\pb{\phi^I(\bx),p_J(\by)}=\delta^I_J\delta(\bx-\by) \ , \quad 
\pb{T(\bx),p_T(\by)}=\delta(\bx-\by) \ . \nonumber \\	
\end{eqnarray}

First of all we calculate Poisson brackets between spatial 
diffeomorphism constraints and we easily get
\begin{equation}
	\pb{\bC_S(X^i),\bC_S(Y^j)}=\bC_S(X^j\partial_j Y^i-Y^j\partial_j X^i) \ . 
\end{equation}
In case of the Poisson bracket between the constraints $\mC$ we firstly  calculate
\begin{eqnarray}
&& \pb{\bC^{matt}(X),
	\bC^{matt}(Y)}=	\pb{\int d^9\bx X \sqrt{\mD+\mH_ih^{ij}\mH_j},
		\int d^9\by Y\sqrt{\mD+\mH_ih^{ij}\mH_j}}=
	\nonumber \\
&&	=\frac{1}{4}
	\int d^9\bx d^9\by \frac{X}{\sqrt{\mD+\mH^ih^{ij}\mH_j}}(\bx)
	\left\{(\mD+\mH_ih^{ij}\mH_j)(\bx), 	(\mD+\mH_ih^{ij}\mH_j)(\by)\right\}\times 
	\nonumber \\
&&	\frac{Y}{\sqrt{\mD+\mH^ih^{ij}\mH_j}}(\by)	=
	\nonumber \\		
&&	=\int d^9\bx (X\partial_iY-Y\partial_iX)h^{ij}\mH_j-\nonumber \\
&&	-\int d^9\bx (X\partial_iY-Y\partial_iX)\frac{1}{\mD+\mH_ih^{ij}\mH_j}
	[(\bA^{-1})^{ij}_S\bA_{jk}^A+(\bA^{-1})^{ij}_A\bA_{jk}^S]\pi^kV^2\det \bA=
	\nonumber \\
&&	=\int d^9\bx (X\partial_iY-Y\partial_iX)h^{ij}\mH_j \ , \nonumber \\
\end{eqnarray}
where we again used 
\begin{eqnarray}
&&	(\bA^{-1})^{ij}_S \bA_{
		jk}^S+(\bA^{-1})^{ij}_A
	\bA_{jk}^A=\delta^i_k \ , \nonumber \\
&&	(\bA^{-1})^{ik}_A\bA_{
		kj}^S+(\bA^{-1})^{ik}_S \bA^A_{kj}=0 \ . 
\end{eqnarray}
If we take into account well known result related to the gravitational part,  \cite{Arnowitt:1962hi}
\begin{equation}
	\pb{\bC^G(X),\bC^G(Y)}=
	\bC_S^G((X\partial_iY-Y\partial_iX)h^{ij})
\end{equation}
and also used the fact that 
\begin{equation}
	\pb{\bC^G(X),\bC^{matt}(Y)}+
	\pb{\bC^{matt}(X),\bC^G(Y)}=0
\end{equation}
we obtain desired result
\begin{equation}
	\pb{\bC(X),\bC(Y)}=\bC_S((X\partial_iY-Y\partial_iX)h^{ij}) \ .
\end{equation}
Finally we calculate Poisson bracket
\begin{eqnarray}
	\pb{\bC_S(X^i),\mC^{matt}(\bx)} \ . \nonumber \\
\end{eqnarray}
Since
\begin{eqnarray}
&&	\pb{\bC_S(X^i),\mD}=-X^m\partial_m\mD-2\partial_m X^m\mD \ , \nonumber \\
&&	\pb{\bC_S(X^i),\mH_i}=-\partial_i X^m\mH_m-\partial_mX^m\mH_i-X^m\partial_m\mH_i \ , 
	\nonumber \\
\end{eqnarray}
we get 
\begin{equation}
	\pb{\bC_S(X^i),\mC^{matt}}=-\partial_m X^m\mC^{matt}-
	X^m\partial_m \mC^{matt} \ 
\end{equation}
that together with 
\begin{equation}
\pb{\bC_S(X^i),\mH^{G}_\bot}=-\partial_m X^m\mH^G_\bot-
X^m\partial_m \mH^G _\bot\ 
\end{equation}
implies the final result 
\begin{equation}
	\pb{\bC_S(X^i),\mC}=-\partial_m X^m\mC-
	X^m\partial_m \mC \ .
\end{equation}
Taking all these results together we obtain  that $\mC\approx 0$ and $\mC_i\approx 0$ are first class constraints. 

Now we are ready to proceed to the application  of Brown-Kucha\v{r} mechanism to the case of unstable D9-brane.

\section{Brown-Kucha\v{r} Mechanism for Unstable D-Brane}
\label{third}
In this section we propose  Brown Kuchar mechanism  for  space-time filling unstable D9-brane coupled to gravity. 
The first step is to separate contribution proportional to $\partial_i T\partial_j T $ in $\bA_{ij}$. To do this we write the matrix  $\bA_{ij}$  as
\begin{equation}
\bA_{ij}=\ba_{ij}+\lambda\partial_i T\partial_j T \ , \quad  \ba_{ij}=h_{ij}+
M_{IJ}\partial_i \phi^I\partial_j \phi^J+\lambda F_{ij} \ . 
\end{equation}
Then we can write
\begin{eqnarray}
\det \bA_{ij}=\det \ba_{ij}\det(\delta^j_k+\lambda\ba^{jm}\partial_m T\partial_j T)
=\det \ba (1+\lambda\ba^{ij}\partial_i T\partial_jT) \ ,
\end{eqnarray}
where $\ba^{ij}$ is matrix inverse to $\ba_{ij}$
\begin{equation}
\ba_{ij}\ba^{jk}=\delta_i^k \ . 
\end{equation}
Now using $\mC_i\approx 0$ we can express $
\partial_iT$ as 
\begin{equation}
\partial_i T= -\frac{1}{p_T}\tmH_i \ , \quad \tmH_i=\mH_i^G+p_I\partial_i\phi+F_{ij}\pi^j
\end{equation}
so that the constraint $\mC\approx 0$ takes the form 
\begin{eqnarray}\label{CfuntpT}
\mC=
 \mH^{G}_\bot+\sqrt{\mS+\frac{1}{\lambda}p_T^2+\mR\frac{\lambda}{p_T^2}}\approx 0  \ , 
\nonumber \\
\end{eqnarray}
where
\begin{eqnarray}
&&\mS\equiv 
p_I M^{IJ}p_J+\frac{1}{\lambda^2}\pi^i\ba^S_{ij}\pi^j
+V^2\det \ba_{ij}+\mH_i^{G}h^{ij}\mH_j^{G} \ , \nonumber \\
&&\mR=\frac{1}{\lambda^2}\pi^i\tmH_i\tmH_j\pi^j+V^2\det \ba_{ij}\ba^{ij}\tmH_i\tmH_j\approx 0 \ .
\nonumber \\
\end{eqnarray}
We can solve the constraint $\mC\approx 0$  for $p_T$ as
\begin{equation}
\frac{1}{\lambda}p^2_T=\frac{((\mH^{G}_\bot)^2-\mS)+
\sqrt{(\mS-(\mH^{G}_\bot)^2)^2-4\mR}}{2}
\end{equation}
so that we have new constraint
\begin{eqnarray}\label{mGdef}
&&\mG \equiv -\frac{1}{\sqrt{\lambda}}p_T+\sqrt{\frac{1}{2}((\mH^{GR})^2-\mS)
	+\sqrt{\frac{1}{4}(\mS-(\mH^{GR})^2)^2
		-\mR}}\equiv
	\nonumber \\
&&	\equiv  -\frac{1}{\sqrt{\lambda}}p_T+H \  
	\approx 0 \ , \nonumber \\
\end{eqnarray}
where we presume that $H$ is positive since we want to identify it
with physical Hamiltonian density so that it is natural to define  Hamiltonian as  
\begin{equation}
\bH=\int d^9\bx H(\bx) \  
\end{equation}
that is positive by definition and which is non-zero. 

It is clear that the analysis presented above is valid for general $T$ which means that
$\bH$ depends on $T$ explicitly. In fact, $\mG\approx 0$ defined above should be the first
class constraint too. We will explicitly calculate algebra of constraints $\mG(\bx),\mG(\by)$ for general 
$T$ in future publication. On the other hand in order to identify $T$ with time 
it is natural to presume that $T$ is large so that we can neglect term proportional to 
$V(T)$ which for large $T$ goes to zero.  Now we show that Hamiltonian density as defined above
Poisson commutes with all constraints. 
\subsection{Calculation of Poisson brackets between Hamiltonian}
In this section we would like to study properties of $\mG$ and $H$ defined in  (\ref{mGdef}).
First of all we would like to calculate Poisson bracket between following objects
\begin{equation}
G(\bx)=(H^G_\bot)^2-H_i^G h^{ij}H_j^G \ , \quad 
\bG(X)=\int d^9\bx XG(\bx) \ , 
\end{equation}
whose importance was firstly stressed in 
\cite{Brown:1994py} and further studied in \cite{Kuchar:1995xn,Markopoulou:1996nh,Thiemann:2006up}.
Following \cite{Brown:1994py} we introduce 
 $F(X)$ as
\begin{equation}
	F(X)=\int d^9\bx X \mH^G_i h^{ij}\mH^G_j \ 
\end{equation}
and calculate 
\begin{eqnarray}
&&	\pb{F(X),F(Y)}=
	4\int d^9\bx d^9\by X\mH^G_i h^{ij}\pb{\mH^G_j(\bx),\mH^G_k(\by)}h^{kl}\mH^G_l Y+
	\nonumber \\
&&	+2\int d^9\bx d^9\by X \mH^G_i\mH^G_j\pb{h^{ij},\mH^G_k(\by)}h^{kl}\mH^G_l(\by)Y(\by)+
	\nonumber \\
&&	2\int d^9\bx d^9\by X\mH^G_i h^{ij}\pb{\mH^G_j(\bx),h^{kl}(\by)}\mH^G_k\mH^G_l(\by)Y(\by)=0
	\nonumber \\
\end{eqnarray}
using the fact that 
\begin{equation}
	\pb{\int d^9\bx X^m\mH_m^G,h_{ij}}=-X^m\partial_m h_{ij}-\partial_i X^m h_{mj}-
	h_{im}\partial_jX^m \ . 
\end{equation}
Then we calculate  
\begin{eqnarray}\label{PBG}
&&\pb{\bG(X),\bG(Y)}=
\pb{\int d^9\bx X(\mH^G_\bot)^2,\int d^9\by Y (\mH^G_\bot)^2}-\nonumber \\
&&-\pb{\int d^9\bx X(\mH^G_\bot)^2,F(Y)}
-\pb{F(X),\int d^9\by 
Y (\mH^G_\bot)^2} \ .  \nonumber \\
\end{eqnarray}
The expression on the first line in (\ref{PBG}) is equal to
\begin{eqnarray}\label{firstline}
&&\pb{\int d^9\bx X(\mH^G_\bot)^2,\int d^9\by Y (\mH^G_\bot)^2}=\nonumber \\
&&=4\int d^9\bx (X\partial_iY-Y\partial_iX)h^{ij}\mH^G_j (\mH^G_\bot)^2
\nonumber \\
\end{eqnarray}
while calculation of the Poisson brackets on the second line leads to 
\begin{eqnarray}\label{secondline}
&&-\pb{\int d^9\bx X(\mH^G_\bot)^2,F(Y)}-
\pb{F(X),\int d^9\by 
	Y (\mH^G_\bot)^2}=
\nonumber \\
&&4\int d^D\bx (Y\partial_iX-X\partial_iY)h^{ij}\mH^G_j (\mH^G_\bot)^2 \ .
\nonumber \\
\end{eqnarray}
where we used the fact that 
\begin{eqnarray}
\pb{\bT_S(X^i),\mH^G_\bot}=
-X^i\partial_i\mH^G_\bot-\partial_m X^m\mH^G_\bot \ . \nonumber \\
\end{eqnarray}
Taking (\ref{firstline}) and (\ref{secondline}) together 
 we reproduce an important result \cite{Brown:1994py}
\begin{equation}
\pb{\bG(X),\bG(Y)}=0 \ .
\end{equation}
Let us now return to the case of space-time filling D9-brane
with the large tachyon field. In this approximation we can neglect term proportional to $V(T)$ and hence  $\mM$ and $\mR$ are equal to
\begin{eqnarray}
&&\mM=(\mH^{G}_\bot)^2-p_IM^{IJ}p_J-\frac{1}{\lambda^2}\pi^i\ba_{ij}^S\pi^j-\mH_i^{G}h^{ij}\mH_j^{G}=
\nonumber \\
&&=G-p_IM^{IJ}p_J-\frac{1}{\lambda^2}\pi^i\ba_{ij}^S\pi^j \ , 
\quad 
\mR=\frac{1}{\lambda^2}\pi^i\tmH_i\tmH_j\pi^j \ , \nonumber \\
&&\ba^S_{ij}=h_{ij}+M_{IJ}\partial_i\phi^I\partial_j\phi^J \ , \quad \tmH_i=\mH_i^G+
p_I\partial_i\phi^I+F_{ij}\pi^j \ . \nonumber \\
\end{eqnarray}
Let us introduce  $\bM(X)=\int d^9\bx X\mM$ 
and  calculate following Poisson bracket 
\begin{eqnarray}
&&\pb{\bM(X),\bM(Y)}=
\pb{\bG(X),\bG(Y)}-\nonumber \\
&&\pb{\bG(X),\int d^9\by Y(p_IM^{IJ}p_J+\frac{1}{\lambda^2}\pi^i\ba_{ij}^S\pi^j)}
-\nonumber \\
&&-\pb{\int d^9\bx X(p_IM^{IJ}p_J+\frac{1}{\lambda^2}\pi^i\ba_{ij}^S\pi^j),
	\bG(Y)}+
\nonumber \\
&&\pb{\int d^9\bx X(p_IM^{IJ}p_J+\frac{1}{\lambda^2}\pi^i\ba_{ij}^S\pi^j),
\int d^9\by Y (p_K M^{KL}p_L+\frac{1}{\lambda^2}\pi^k\ba_{kl}^S\pi^l)}
\nonumber \\
&&=4\int d^D\bx(X\partial_iY-Y\partial_iX)\frac{1}{\lambda^2}\pi^i\tmH_j\pi^j \ , 
\nonumber \\ 
\end{eqnarray}
%
where we used the fact that 
\begin{eqnarray}
	\pi^i\tmH_i=
\pi^i(\mH_i^G+F_{ij}\pi^j+p_I\partial_i\phi^I)=
\pi^i(\mH_i^G+p_I\partial_i\phi^I)	\ . 
\nonumber \\
\end{eqnarray}
As the final step we  
 define $\bH(X)$ as
\begin{equation}
\bH(X)=\int d^9\bx XH(\bx) \ , 
\end{equation}
where 
\begin{equation}
H=
	\sqrt{\frac{\mM}{2}+\sqrt{\frac{1}{4}\mM^2-\mR}} \ , 
\end{equation}		
and where $X$ is arbitrary test function. Then we can calculate
following  Poisson brackets 
\begin{eqnarray}
&&	\pb{\bH(X),\bH(Y)}
=\nonumber \\
&&	=\frac{1}{16}\int d^9\bx d^9
\by
\frac{X(\bx)}{\sqrt{\frac{1}{2}\mM+\sqrt{\frac{1}{4}\mM^2-\mR}}}
\pb{\mM(\bx),\mM(\by)}\frac{Y(\by)}{\sqrt{\frac{1}{2}\mM+\sqrt{\frac{1}{4}\mM^2-\mR}}}+
\nonumber \\
&&+	\frac{1}{16}\int d^9\bx d^9
\by
\frac{X(\bx)}{\sqrt{\frac{1}{2}\mM+\sqrt{\frac{1}{4}\mM^2-\mR}}}\times\nonumber \\
&&\pb{\mM(\bx),(\frac{1}{4}\mM^2-\mR)(\by)}\frac{Y(\by)}{\sqrt{\frac{1}{2}\mM+\sqrt{\frac{1}{4}\mM^2-\mR}}\sqrt{\frac{1}{4}\mM^2-\mR}}+
\nonumber \\
&&+	\frac{1}{16}\int d^9\bx d^9
\by
\frac{X(\bx)}{\sqrt{\frac{1}{2}\mM+\sqrt{\frac{1}{4}\mM^2-\mR}}\sqrt{\frac{1}{4}\mM^2-\mR}}\times\nonumber \\
&&\pb{(\frac{1}{4}\mM^2-\mR)(\bx),\mM(\by)}\frac{Y(\by)}{\sqrt{\frac{1}{2}\mM+\sqrt{\frac{1}{4}\mM^2-\mR}}}+
\nonumber \\
&&+	\frac{1}{16}\int d^9\bx d^9
\by
\frac{X(\bx)}{\sqrt{\frac{1}{2}\mM+\sqrt{\frac{1}{4}\mM^2-\mR}}\sqrt{\frac{1}{4}\mM^2-\mR}}\times\nonumber \\
&&\pb{(\frac{1}{4}\mM^2-\mR)(\bx),(\frac{1}{4}\mM^2-\mR)(\by)}\frac{Y(\by)}{\sqrt{\frac{1}{2}\mM+\sqrt{\frac{1}{4}\mM^2-\mR}}\sqrt{\frac{1}{4}\mM^2-\mR}}=\nonumber \\
&&=\frac{1}{4\lambda^2}\int d^9\bx
(X\partial_iY-Y\partial_iX)\pi^i\tmH_j\pi^j
\frac{1}{\frac{1}{2}\mM+\sqrt{\frac{1}{4}\mM^2-\mR}}-\nonumber \\
&&-\frac{1}{4\lambda^2}\int d^9\bx (X\partial_mY-Y\partial_mX)\pi^m \pi^j
\tmH_j 
\frac{(\frac{1}{4}\mM^2-\mR)}
{(\frac{1}{2}\mM+\sqrt{\frac{1}{4}\mM^2-\mR}}
\frac{1}{\frac{1}{4}\mM^2-\mR}=0
\nonumber \\
	\end{eqnarray}
using the fact that 
\begin{eqnarray}
&&\pb{\int d^9
	\bx X\mM(\bx),\int d^9\by 
	Y \mR(\by)}=
-\frac{4}{\lambda^2}\int d^9\bx (Y\partial_mX-X\partial_mY)\pi^m\tmH_j\pi^j\mM
\nonumber \\
\end{eqnarray}	
and also
\begin{eqnarray}
&&\pb{\int d^9\bx X^i\tmH_i,\mM(\by)}=-X^i\partial_i\mM-2\partial_m X^m\mM \ , 
\nonumber \\
&&-\pb{\int d^9\bx X\mM^2,\int d^9\by Y \mR}-
\pb{\int d^9\bx X\mR,\int d^9\by Y \mM^2}=
\nonumber \\
&&=-\frac{8}{\lambda^2}\int d^9\bx(\partial_m YX-Y\partial_mX)\pi^m\pi^k\tmH_k \mM^2 \ . 
\nonumber \\
\end{eqnarray}	
Finally we used following result 
\begin{eqnarray}
&&	\pb{\int d^9\bx X\mR,
		\int d^9\by Y \mR}
%
	=\frac{4}{\lambda^4}\int d^9\bx (X\partial_iY-Y\partial_iX)\pi^i \tmH_j\pi^j(\pi^m\tmH_m)^2 \nonumber \\
\end{eqnarray}
The result that $\pb{\bH(X),\bH(Y)}=0$ has an important consequence. Since $\mH(\bx)$ does not depend on $T$ for large $T$  we have $\pb{p_T(\bx),\mH(\by)}=0$ and hence the constraints $\mG(\bx)\approx 0$
defined in (\ref{mGdef}) have vanishing Poisson brackets 
\begin{equation}
	\pb{\mG(\bx),\mG(\by)}=0 \ 
\end{equation}
which is generalization of the result derived in \cite{Brown:1994py} to the case of the space-time filling 
unstable D9-brane in the limit of large tachyon field $T$.

Finally we will calculate Poisson bracket between $\bC_S(X^i)$ and $\mG$. Using
\begin{equation}
	\pb{\bC_S(X^i),H}=-X^m\partial_m H-\partial_m X^m H
\end{equation}
we obtain 
\begin{equation}
	\pb{\bC_S(X^i),\mG}=-X^m\partial_m\mG-\partial_m X^m\mG \ 
\end{equation}
that shows that $\mG$ transforms as tensor density.Then  we easily find that 
 $\pb{\bC_S(X^i),\bH}=0$.

Now we have all ingredients necessary for deparametrization of gravity coupled to the tachyon, following 
\cite{Thiemann:2006up}.

\section{Deparametrization of General Relativity Coupled to Tachyon}\label{fourth}
In previous section we showed that $\bH$ Poisson commutes with 
all constraints which is necessary condition for deparametrization of gravity.   Following 
\cite{Thiemann:2006up} we  define $\bH_\tau$ as
\begin{equation}
	\bH_\tau=\int d^9\bx[\tau-\sqrt{\lambda}T(\bx)]H(\bx) \ ,
\end{equation}
where $\tau$ has physical dimension of length which is appropriate for time variable. Note that then $\bH_\tau$ is dimensionless due to the fact that $H$ is proportional to $l_s^{10}$.

Let $f$ is spatial diffeomorphism invariant quantity that does not depend on $T$ and $p_T$. 
Then we define observable $O_f(\tau)$ by following prescription
\begin{equation}\label{defOf}
	O_f(\tau)=\sum_{n=0}^\infty
	\frac{1}{n!} \pb{\bH_\tau,f}_{(n)} \ , 
\end{equation}
where multiple Poisson brackets are defined iteratively 
\begin{eqnarray}
&&	\pb{\bH_\tau,f}_{(0)}=f\ , 
	\quad \pb{\bH_\tau,f}_{(1)}=\pb{\bH_\tau,\pb{\bH_\tau,f}_{(0)}}=
	\pb{\bH_\tau,f} \ , \nonumber \\
&&	\pb{\bH_\tau,f}_{(n+1)}=\pb{\bH_\tau,\pb{\bH_\tau,f}_{(n)}} \ . 
\end{eqnarray}
Let us now calculate following derivative with respect to $\tau$
\begin{eqnarray}\label{taupn}
&&	\frac{d}{d\tau}\pb{\bH_\tau,f}_{(n)}=
\pb{\bH,\pb{\bH_\tau,f}_{(n-1)}}+
	\pb{\bH_\tau,\frac{d}{d\tau}\pb{\bH_\tau,f}_{(n-1)}}=
	\nonumber \\
&&=\pb{\bH,\pb{\bH_\tau,f}_{(n-1)}}+
\pb{\bH_\tau,\pb{\bH,\pb{\bH_\tau,f}_{(n-2)}}}+
\pb{\bH_\tau,\pb{\bH_\tau,\frac{d}{d\tau}	\pb{\bH_\tau,f}_{(n-2)}}}=\nonumber \\	
&&=2\pb{\bH,\pb{\bH_\tau,f}_{(n-1)}}+	\pb{\bH_\tau,\pb{\bH_\tau,\frac{d}{d\tau}	\pb{\bH_\tau,f}_{(n-2)}}}=\nonumber \\
&&\quad \vdots \nonumber \\
&&n\pb{\bH,\pb{\bH_\tau,f}_{(n-1)}} \ , \nonumber \\
\end{eqnarray}
where we used 
 Jacobi identity of Poisson brackets that hold for any phase space functions $A,B,C$ 
\begin{equation}\label{Jaciden}
	\pb{A,\pb{B,C}}+\pb{B,\pb{C,A}}+\pb{C,\pb{A,B}}=0 
\end{equation}
and also the fact that 
\begin{equation}
	\pb{\bH,\bH_\tau}=0 \
\end{equation}
so that 
\begin{equation}
	\pb{\bH_\tau,\pb{\bH,\pb{\bH_\tau,f}_{(n-2)}}}=
	\pb{\bH,\pb{\bH_\tau,\pb{\bH_\tau,f}_{(n-2)}}} \ . 
\end{equation}
Then using (\ref{taupn}) we can easily determine $\tau-$ derivative of the function $O_f$ defined in 
(\ref{defOf})
\begin{eqnarray}
&&	\frac{d}{d\tau}Q_f(\tau)=
	\sum_{n=0}^\infty\frac{1}{n!}\frac{d}{d\tau}
	\pb{\bH_\tau,f}_{(n)}=\nonumber \\
&&=\sum_{n=1}^\infty
\pb{\bH,\pb{\bH_\tau,f}_{(n-1)}}=\pb{\bH,O_f}
\nonumber \\	
\end{eqnarray}
which has the form of the Hamiltonian equation that expresses true evolution with respect to the time parameter $\tau$ generated by Hamiltonian $\bH$.
%
%
%
Further, since we claim that $O_f$ is Dirac's observable it should also have vanishing Poisson brackets with constraints
$\mG\approx 0 , \mC_i\approx 0$. In the following paragraphs we show that it is really true.

\subsection{Hamiltonian Constraint}
Let us define 
 smeared form of the  constraint $\mG$ as
\begin{equation}
	\bG(M)=\int d^9\by M(\by)\mG(\by) \ , \quad \mG=-\frac{1}{\sqrt{\lambda}}p_T+H \ . 
\end{equation}
First of all we have
\begin{eqnarray}
	\pb{\bG(M),\bH_\tau}=-\bH(M) \  
\nonumber \\
\end{eqnarray}
using the fact that $\pb{\bH(M),\bH}=0$. Before we proceed further 
 let us define $\bP_{(n)}$ as 
\begin{equation}
	\bP_{(n)}\equiv \pb{\bH_\tau,f}_{(n)} \ 
\end{equation}
in order to make notation more tractable. Then we calculate Poisson 
bracket between $\bG(M)$ and $\bP_{(n)}$
\begin{eqnarray}\label{bGham}
&&\pb{\bG(M),\bP_{(n)}}=
\pb{\bG(M),\pb{\bH_\tau,\bP_{(n-1)}}}=	
	\nonumber \\
&&=\pb{\bH_\tau,\pb{\bG(M),\bP_{(n-1)}}}-
\pb{\bH(M),\bP_{(n-1)}}=\nonumber \\
&&=\pb{\bH_\tau,\pb{\bH_\tau,\pb{\bG(M),\bP_{(n-2)}}}}-2
\pb{\bH(M),\bP_{(n-1)}}=
\nonumber \\
&&=\quad \vdots \nonumber \\
&&=\pb{\bH_\tau,\pb{\bH_\tau,\dots 
\pb{\bG(M),f}}\dots}-n
\pb{\bH(M),\bP_{(n-1)}} \ , 
\nonumber \\			
\end{eqnarray}
where we again used 
(\ref{Jaciden}) and the fact that
$\pb{\bH(M),\bH_\tau}=0$  which   implies 
\begin{equation}
	\pb{\bH(M),\pb{\bH_\tau,\dots}}=\pb{\bH_\tau,\pb{\bH(M),\dots}} \ . 
\end{equation}
But 
\begin{equation}
	\pb{\bG(M),f}=
	\pb{\bH(M),f} 
\end{equation}
since phase space function $f$ does not depend on $T$ and $p_T$ by definition. Then (\ref{bGham}) can be rewritten into the final form
\begin{eqnarray}\label{bGhamfinal}
&&\pb{\bG(M),\bP_{(n)}}=
\pb{\bH_\tau,\pb{\bH_\tau,\dots,
		\bH(M),f}\dots}
-\pb{\bH(M),\bP_{(n-1)}}=
\nonumber \\
&&=\pb{\bH(M),\bP_{(n)}}-n\pb{\bH(M),\bP_{(n-1)}}
	\nonumber \\
\end{eqnarray}
which is central formula for what follows.  More precisely, let us now calculate Poisson bracket between $\bG(M)$ and $O_f$ 
\begin{eqnarray}
&&	\pb{\bG(M),O_f}=
\pb{\bG(M),f}+\sum_{n=1}^\infty\frac{1}{n!}
\pb{\bG(M),\bP_{(n)}}=\nonumber \\
&&=\pb{\bH(M),f}+
\sum_{n=1}^\infty \frac{1}{n!}
\pb{\bH(M),\bP_{(n)}}-
	\sum_{n=1}^\infty
\frac{1}{n!} n\pb{\bH(M),\bP_{(n-1)}}=
\nonumber \\
&&=\sum_{n=1}^\infty \frac{1}{n!}
\pb{\bH(M),\bP_{(n)}}-
				\sum_{n=2}^\infty
			\frac{1}{(n-1)!} \pb{\bH(M),\bP_{(n-1)}}= \nonumber \\
&&	\sum_{n=1}^\infty
\frac{1}{n!} n\pb{\bH(M),\bP_{(n)}}-
	\sum_{l=1}^\infty
\frac{1}{l!} \pb{\bH(M),\bP_{(l)}}=0 \ , 
\nonumber \\
\end{eqnarray}
where in the final step we introduced variable $l=n-1$. In other words we proved that variable  $O_f$ Poisson commutes with 
Hamiltonian constraint.
\subsection{Spatial Diffeomorphism Constraint}
Finally we will check that $O_f$ Poisson commutes with $\bC(M^i)$. Since $f$ is spatially invariant function by definition we have 
\begin{equation}\label{spatO}
	\pb{\bC(M^i),f}=0 \ .
\end{equation}
Further, in case of $\bH_\tau$ we have 
\begin{eqnarray}\label{spatH}
&&	\pb{\bC(M^i),\bH_\tau}=
\int d^9\bx	\pb{\bC(M^i),[\tau-\sqrt{\lambda}T(\bx)]}H(\bx)+
\nonumber \\
&&+\int d^9\bx [\tau-\sqrt{\lambda}T(\bx)]\pb{\bC(M^i),H(\bx)}=
	\nonumber \\
&&=\sqrt{\lambda}\int d^9\bx M^i\partial_iT H(\bx)+
\int d^9\bx [\tau-\sqrt{\lambda}T(\bx)](-M^i\partial_i\mH-\partial_mM^m\mH)=0
 \ . 
\nonumber \\
\end{eqnarray}
Then if we combine (\ref{spatO}) with (\ref{spatH}) we obtain final result 
\begin{equation}
	\pb{\bC(M^i),O_f}=0 \ . 
\end{equation}
that proves an invariance of phase space functions $O_f$ under spatial diffeomorphism.

In conclusion, we performed  construction 
of Dirac observables following very nice analysis presented in 
\cite{Thiemann:2006up}. These observables evolve according
to Hamiltonian equations with respect to $\tau$
 and which have vanishing Poisson brackets with Hamiltonian and spatial diffeomorphism 
 constraints which mean that they are strong Dirac observables. 
 Our analysis is different from  the treatment presented in 
\cite{Thiemann:2006up}. Further, it was used for the first time in case  
of space-time feeling non-BPS D9-brane. It also  demonstrates that  open string tachyon at least for its large value,  is natural time variable which is in the agreement with A. Sen's proposal
\cite{Sen:2023qya,Sen:2002qa}.

{\bf Acknowledgement:}

This work  is supported by the grant “Dualitites and higher order derivatives” (GA23-06498S) from the Czech Science Foundation (GACR).

\end{document}